\documentclass[aps,prb,twocolumn,showpacs]{revtex4}
\usepackage{amssymb}
\usepackage{graphicx}
\usepackage{amsmath}
\usepackage{amsfonts}

\begin{document}

\title{Spin Liquid States at the Vicinity of Metal-Insulator Transition}
\date{\today}
\author{Yi Zhou$^{1}$ and Tai-Kai Ng$^2$}
\affiliation{Department of Physics and Zhejiang Institute of Modern Physics, Zhejiang
University, Hangzhou, 310027, P. R. China$^1$\\
Department of Physics, Hong Kong University of Science and Technology, Clear
Water Bay Road, Kowloon, Hong Kong, China$^2$}

\begin{abstract}
We study in this paper quantum spin liquid states (QSLs) at the vicinity of
metal-insulator transition. Assuming that the low energy excitations in the
QSLs are labeled by ``spinon" occupation numbers with the same Fermi surface
structure as in the corresponding metal (Fermi-liquid) side, we propose a
phenomenological Landau-like low energy theory for the QSLs and show that
the usual $U(1)$ QSLs is a representative member of this class of spin
liquids. Based on our effective low energy theory, an alternative picture to
the Brinkman-Rice picture of Mott metal-insulator transition is proposed.
The charge, spin and thermal responses of QSLs are discussed under such a
phenomenology.
\end{abstract}

\maketitle

%75.10.Kt Quantum spin liquids, valence bond phases and related phenomena
%71.10.-w Theories and models of many-electron systems
%71.10.Ay Fermi-liquid theory and other phenomenological models
%71.30.+h Metal-insulator transitions and other electronic transitions

\section{Introduction}

Quantum spin liquid states (QSLs) in dimensions $d>1$ has been a long sought
dream in condensed matter physics which has not been confirmed in realistic
materials until rather recently.\cite{Lee08} These states are electronic
Mott insulators that are not magnetically ordered down to the lowest
temperature due to strong quantum mechanical fluctuations of spins and/or
frustrated interaction. Various exotic properties have been proposed to
exist in QSLs. For instance, charge neutral and spin-$\frac{1}{2}$ mobile
objects, spinons, were proposed to emerge in such electronic states at low
temperature accompanied by different kinds of (emerging) gauge fields. The
spinons may be gapped or gapless and may obey either boson or fermion
statistics. These new particles and gauge fields which characterize low
energy behaviors of the system do not appear in the parent Hamiltonian and
``emerge" as a result of strong correlation.

In the past few years, several experimental candidates for QSLs have been
discovered that support the existence of gapless fermionic spinon
excitations. The best studied example is a family of organic compounds $%
\kappa -$(ET)$_{2}$Cu$_{2}$(CN)$_{3}$ (ET)\cite{Kanoda03} and Pd(dmit)$_{2}$%
(EtMe$_{3}$Sb) (dmit salts)\cite{Itou08}. Both materials are Mott insulators
in proximity to the metal-insulator transition because they become
superconductor (ET) or metal (dmit) under modest pressure. Despite the large
magnetic exchange $J\approx 250$~K observed in these systems, there is no
experimental indication of long range magnetic ordering down to temperature $%
\sim 30~mK$. Linear temperature dependence of the specific heat and
Pauli-like spin susceptibility were found in both materials at low
temperature suggesting that the low energy excitations are spin-1/2 fermions
with a Fermi surface.\cite{SYamashita,Matsuda12} This Fermi liquid-like
behavior is further supported by their Wilson ratios which are close to one.
The thermal conductivity experiments on the ET salts found a large
contribution to $\kappa $ beside phonons with $\kappa /T$ much reduced below
0.3~K,\cite{MYamashita09} while $\kappa /T$ approaches to a constant down to
the lowest temperature in dmit salts.\cite{MYamashita10} All these
experimental observations point to the scenario that the low lying
excitations in these Mott insulators are mobile fermionic particles
(spinons) that form a Fermi surface like their parent electrons ($U(1)$ spin
liquid state). Besides ET and dmit-salts, the Kagome compound ZnCu$_{3}$(OH)$%
_{6}$Cl$_{2}$, the three dimensional hyper-Kagome material Na$_{4}$Ir$_{3}$O$%
_{8}$ and the newly discovered triangular compound Ba$_{3}$CuSb$_{2}$O$_{9}$
are also considered to be candidates for QSLs with gapless excitations.\cite%
{Kagome07,Takagi07,HDZhou2011}

Several experiments were proposed to probe mobile spinons in the $U(1)$ spin
liquid state. For example, giant magnetoresistance like experiment was
designed to measure mobile spinons through oscillatory coupling between two
ferromagnets via a quantum spin liquid spacer.\cite{Norman2009} The thermal
Hall effect in insulating quantum magnets was proposed as an example of
thermal transport of spinons, where different responses were used to
distinguish between magnon- and spinon- transports.\cite{KNL2010} The spinon
life time and mass as well as gauge fluctuations can be measured through
sound attenuation experiment.\cite{YZhou2011} Despite all these proposals, a
generic method to compare theoretical prediction of QSLs to experimental
data is still missing at the phenomenological level.

The purpose of this paper is to build a generic phenomenological theory for
spin liquids with (large) Fermi surfaces. Starting from the fact that these
QSLs are electrical insulators but good thermal conductors, we propose a
unified Fermi liquid type effective theory that describes both Fermi liquids
and QSLs with large Fermi surfaces. The theoretical framework allows us to
compute thermodynamics, transport and electromagnetic response of QSLs
coherently and compare the results with experiments.

The paper is organized as follows. In Section II, we discuss our formulation
of Landau's Fermi liquid type effective theory for both Fermi liquid and
spin liquid states. In Section III, we discuss electromagnetic response
where we calculate both AC conductivity and dielectric function for the
QSLs. In Section IV, the renormalization of thermodynamics quantities are
discussed. In Section V, we discuss the transport equation for
quasiparticles where the scattering amplitude and thermal conductivity are
computed base on the transport equation. In Section VI, we point out the
connection between our theory and $U(1)$ gauge theory and propose an
alternative picture from Brinkman-Rice's for the Mott metal-insulator
transition. The important implication of the Pomeranchuk instability is
pointed out. Section VII is devoted to the summary.

\section{Phenomenological theory: Landau's Fermi liquid type effective theory%
}

In a Fermi liquid, when electron-electron interactions are adiabatically
turned on, the low energy excited states of interacting N-electron systems
evolve in a continuous way, and therefore remain in one-to-one
correspondence with the states of noninteracting N-electron systems. The
same labeling scheme through fermion occupation number is assumed in the
theories of $U(1)$ spin liquid state. we shall make the same assumption here
when we consider general fermionic QSLs with finite spinon Fermi surfaces,
although the one-to-one correspondence with the states of noninteracting
electron systems is \emph{not} protected by adiabaticity and should be
viewed as an ansatz.

With this assumption the low energy excitations in the QSLs with finite
spinon Fermi surfaces are labeled by the same occupation numbers as free
fermions. The difference between Fermi liquids and QSLs is that the
excitations in Fermi liquid are quasi-particles that carry both charges and
spins, whereas the excitations in QSLs are expected to carry only spins. In
particular, DC charge transport exists in Fermi liquid states, but vanishes
in QSLs (insulators). Meanwhile, there exist large electronic contribution
to thermal conductivity at low temperature in these insulating states
because of mobile spin excitations. These two phenomena provide additional
criteria to specifying the Fermi liquid type effective theory for QSLs.

The assumption that the low energy excitations in these QSLs are labeled by
the same occupation numbers as free fermions suggests that the excitation
energy $\Delta E=E-E_{G}$ for these states are also given by a Landau-type
expression\cite{Pines,Baym} 
\begin{equation}
\Delta E=\sum_{\mathrm{p\sigma }}\xi _{\mathrm{p}}\delta n_{\mathrm{p\sigma }%
}+\frac{1}{2}\sum_{\mathrm{pp}^{\prime }\sigma \sigma ^{\prime }}f_{\mathrm{%
pp}^{\prime }}^{\sigma \sigma ^{\prime }}\delta n_{\mathrm{p}\sigma }\delta
n_{\mathrm{p}^{\prime }\sigma ^{\prime }}+O(\delta n^{3}),  \label{F}
\end{equation}%
where $\xi _{\mathrm{p}}=\frac{\mathrm{p}^{2}}{2m^{\ast }}-\mu $ is the
(single) spinon energy measured from the chemical potential $\mu $, $m^{\ast
}$ is the spinon effective mass and $\sigma $ and $\sigma ^{\prime }$ are
spin indices. $\delta n_{\mathrm{\ p\sigma }}=n_{\mathrm{p\sigma }}-n_{%
\mathrm{p\sigma }}^{0}$ measures the departure of the spinon distribution
function from the ground state distribution $n_{\mathrm{p}}^{0}=\theta (-\xi
_{\mathrm{p}})$. $f_{\mathrm{pp}^{\prime }}^{\sigma \sigma ^{\prime }}$ is
the interaction energy between excited spinons. A spherical, rotational
invariant Fermi surface is assumed here for simplicity. In this case we may
write $f_{\mathrm{pp}^{\prime }}^{\sigma \sigma ^{\prime }}$ in terms of
spin symmetric and spin antisymmetric components $f_{\mathrm{pp}^{\prime
}}^{\sigma \sigma ^{\prime }}=f_{\mathrm{pp}^{\prime }}^{s}\delta _{\sigma
\sigma ^{\prime }}+f_{\mathrm{pp}^{\prime }}^{a}\sigma \sigma ^{\prime }$.
For isotropic systems, $f_{\mathrm{pp}^{\prime }}^{s(a)}$ depends only on
the angle $\theta $ between $\mathrm{p}$ and $\mathrm{p}^{\prime }$ and we
can expand $f_{\mathrm{pp}^{\prime }}^{s(a)}=\sum_{l=0}^{\infty
}f_{l}^{s(a)}P_{l}(\cos \theta )$ at 3D and $f_{\mathrm{pp}^{\prime
}}^{s(a)}=\sum_{l=0}^{\infty }f_{l}^{s(a)}\cos (l\theta )$ at 2D, where $%
P_{l}$'s are Legendre polynomials. The Landau parameters, defined by 
\begin{equation*}
F_{l}^{s(a)}=N(0)f_{l}^{s(a)},
\end{equation*}%
provide a dimensionless measures of the strengths of the interactions
between spinons on the Fermi surface, where $N(0)$ is the Fermi surface
density of states. The low temperature properties of the QSLs are completely
determined by the spinon mass $m^{\ast }$ and the interaction $f_{\mathrm{pp}%
^{\prime }}^{\sigma \sigma ^{\prime }}$ (or $F_{l}^{s(a)}$) as in Fermi
liquid theory.

Notice that the energy functional $\Delta E$ for our QSLs is so far
identical to that for Fermi liquids. To describe QSLs, additional conditions
have to be imposed to ensure that the excitations in the effective low
energy theory carry zero charge. We propose and shall demonstrate in the
following that the QSLs distinguish themselves from Fermi liquids by having
a strong constraint on the Landau parameters $F_1^s$.

We start with the observation that the charge current $\mathbf{J}$ carried
by quasi-particles in Fermi liquid theory (and in QSLs) is given by 
\begin{subequations}
\begin{equation}
\mathbf{J}=\frac{m}{m^{\ast }}(1+\frac{F_{1}^{s}}{d})\mathbf{J}^{(0)},
\label{Je}
\end{equation}%
where $\mathbf{J}^{(0)}$ is the charge current carried by the corresponding
non-interacting fermions and $d$ is the dimension. (See Appendix for the
derivation of Eq.(\ref{Je}).) For translational invariant systems, the
charge current carried by quasi-particles is not renormalized and $\frac{%
m^{\ast }}{m}=1+\frac{F_{1}^{s}}{d}$\cite{Baym}. However, this is in general
not valid for electrons in crystals where Galilean invariance is lost. In
this case $\frac{m^{\ast }}{m}\neq 1+\frac{F_{1}^{s}}{d}$ and the charge
current carries by quasi-particles is renormalized by quasi-particle
interaction. On the other hand, the thermal current $\mathbf{J}_{Q}$ is only
renormalized by the effective mass in Fermi liquid theory, 
\begin{equation}
\mathbf{J}_{Q}=\frac{m}{m^{\ast }}\mathbf{J}_{Q}^{(0)},  \label{JQ}
\end{equation}%
where $\mathbf{J}_{Q}^{(0)}$ is the corresponding thermal current carried by
non-interacting electrons. (See Appendix for details.) Thus, in the special
case $1+F_{1}^{s}/d\rightarrow 0$ while $\frac{m^{\ast }}{m}$ remaining
finite, $\mathbf{J}\rightarrow 0$ and $\mathbf{J}_{Q}\neq 0$ suggesting that
the electronic system is in a special state where spin-1/2 quasi-particles
do not carry charge due to interaction but they still carry entropy (i.e.
electric insulating but thermal conducting). This is exactly what we expect
for spinons in QSLs. We note that it is crucial that $F_{1}^{s}$ is
independent of $\frac{m^{\ast }}{m}$ for this mechanism to work.

The charge carried by the quasiparticles (building blocks of our effective
theory) and elementary excitations should be distinguished carefully in our
theory. The building blocks (described by $\delta n_{p\sigma}$'s) are
chargeful quasiparticles, whereas elementary excitations are eigenstates of
Landau's transport equation with charge renormalized by $1+F_{1}^{s}/d$ and
becomes zero in the limit $\mathrm{q}=0$ and $\omega =0$. In general, charge
degrees of freedom are recovered at finite \textrm{q} and $\omega $ as we
shall see when we study electromagnetic responses of QSLs in next section.

\section{Electromagnetic responses}

To put our argument in a more quantitative framework we study the
electromagnetic responses of a Fermi liquid system with $1+F_{1}^{s}/d%
\rightarrow 0$. The charge and (transverse) current response functions are
given by the standard Fermi liquid forms\cite{Ng1,Pines,Leggett} 
\end{subequations}
\begin{equation}
\chi _{d}(\mathrm{q},\omega )=\frac{\chi _{0d}(\mathrm{q},\omega )}{1-\left(
F_{0}^{s}+\frac{F_{1}^{s}(\mathrm{q},\omega )}{d+F_{1}^{s}(\mathrm{q},\omega
)}\frac{\omega ^{2}}{q^{2}}\right) \frac{\chi _{0d}(\mathrm{q},\omega )}{N(0)%
}},  \label{den}
\end{equation}%
and 
\begin{equation}
\chi _{t}(\mathrm{q},\omega )=\frac{\chi _{0t}(\mathrm{q},\omega )}{1-\frac{%
F_{1}^{s}(\mathrm{q},\omega )}{d+F_{1}^{s}(\mathrm{q},\omega )}\frac{\chi
_{0t}(\mathrm{q},\omega )}{N(0)}},  \label{current}
\end{equation}%
where $\chi _{0d}(\mathrm{q},\omega )$ and $\chi _{0t}(\mathrm{q},\omega )$
are the density-density and (transverse) current-current response functions
for a Fermi gas with effective mass $m^{\ast }$ but without Landau
interactions, respectively. The longitudinal current-current response
function $\chi _{l}$ is related to $\chi _{d}$ through 
\begin{equation*}
\chi _{d}(\mathrm{q},\omega )=(q^{2}/\omega ^{2})\chi _{l}(\mathrm{q},\omega
)
\end{equation*}
and the ac conductivity $\sigma _{l(t)}$ is related to $\chi _{l(t)}$ by 
\begin{equation*}
\sigma _{l(t)}(\mathrm{q},\omega )=e^{2}\chi _{l(t)}(\mathrm{q},\omega
)/i\omega ,
\end{equation*}
where $\mathrm{q}=|\vec{q}|$. In the singular limit $1+F_{1}^{s}/d%
\rightarrow 0$, it is clear that higher order $\mathrm{q},\omega $-dependent
terms should be included in the Landau interaction to obtain finite results.
Expanding at small $\mathrm{q}$ and $\omega $, we obtain 
\begin{equation}
\frac{1+F_{1}^{s}(\mathrm{q},\omega )/d}{N(0)}\sim \alpha -\beta \omega
^{2}+\gamma _{t}q_{t}^{2}+\gamma _{l}q_{l}^{2},  \label{f1}
\end{equation}%
where $q_{t}\sim \nabla \times $ and $q_{l}\sim \nabla $ are associated the
transverse (curl) and longitudinal (gradient) parts of the small $\vec{q}$
expansion. $\alpha =0$ in the QSLs. Putting this into the charge response
function $\chi _{d}$, we find that to ensure that the system is in an
incompressible (insulator) state, we must have $\gamma _{l}=0$. The other
possibility $F_{0}^{s}\rightarrow \infty $ implies complete vanishing of
charge responses in the insulating state.

With this parametrization we obtain for the ac conductivity at small $\omega 
$, 
\begin{equation}
\sigma (\omega )=\frac{\omega \sigma _{0}\left( \omega \right) }{\omega
-i\sigma _{0}\left( \omega \right) /\beta e^{2}},  \label{acc}
\end{equation}%
where $\sigma _{0}(\omega )=e^{2}\chi _{0t}(0,\omega )/(i\omega )=e^{2}\chi
_{0l}(0,\omega )/(i\omega )$. The last equality is valid as long as $%
F_{0}^{s}$ is finite. Eq.\ (\ref{acc}) was first obtained in the $U(1)$
gauge theory approach to spin liquid in a slightly different form\cite{nglee}
and predicts power law conductivity Re$[\sigma (\omega )]\sim \omega
^{3.33}(\omega ^{2})$ (at 2D) at frequency regime $\omega >(<)(1/\tau
_{0},k_{B}T/\hbar )$, where $\tau _{0}$ is the elastic scattering time\cite%
{nglee}. The dielectric function is given at small $\mathrm{q},\omega $ by 
\begin{equation}
\varepsilon (\mathrm{q},\omega )=1-\frac{4\pi e^{2}}{q^{2}}\chi _{d}(\mathrm{%
q},\omega )\sim 1+4\pi \beta e^{2}+O(q^{2}),  \label{dielectric}
\end{equation}%
also in agreement with the result obtained in $U(1)$ gauge theory\cite{nglee}%
.

\section{Thermodynamics}

Our picture of QSLs has several immediate experimental consequences. We
shall discuss theromadynamics of QSLs in this section and leave transport
properties in next section.

One of important experimental evidences supporting the existence of gapless
fermionic spinon exciations is linearly temperature dependent specific heat.
The finite specific heat ratio indicates finite density of states at Fermi
level. In Fermi liquid theory, the specific heat ratio $\gamma $ is
renormalized by the effective mass $m^{\ast }$ through the density of states%
\cite{Pines}, namely, 
\begin{equation*}
\gamma =\frac{C_{V}}{T}=\frac{m^{\ast }}{m}\frac{C_{V}^{(0)}}{T}=\frac{%
m^{\ast }}{m}\gamma ^{(0)},
\end{equation*}%
where $C_{V}^{(0)}$ and $\gamma ^{(0)}$ are the specific heat and the
specific heat ratio for the corresponding non-interacting electron gas
respectively. Since $m^{\ast }/m$ remains finite in QSL phase, $\gamma $ is
predicted to be finite at the QSLs.

The spin susceptibility $\chi _{P}$ will be renormalized by the spin
antisymmetric Landau parameter $F_{0}^{a}$ as well as the effective mass $%
m^{\ast }$\cite{Pines}. It gives rise to%
\begin{equation*}
\chi _{P}=\frac{m^{\ast }}{m}\frac{1}{1+F_{0}^{a}}\chi _{P}^{(0)},
\end{equation*}%
where $\chi _{P}^{(0)}$ is the Pauli susceptibility for the corresponding
non-interacting electron gas. It is clear that magnetic susceptibility $\chi
_{P}$ is also non-singular in the QSLs.

Combining the specific heat ratio and spin susceptibility together, we find
that the Wilson ratio 
\begin{equation*}
R=\frac{4\pi ^{2}k_{B}^{2}\chi _{P}}{3(g\mu _{B})^{2}\gamma }\sim
(1+F_{0}^{a})^{-1}
\end{equation*}%
is generally of order $O(1)$, which is one of the experimental observations
in QSLs.

The QSLs have zero compressibility as can be seen from the dielectric
function\ (\ref{dielectric}).

\section{Transport properties}

The transport properties in the QSLs can be computed using the Landau
transport equation. We expect that the transport life-times will be
dominated by scattering in the current-current channel which is the most
singular scattering channel in the limit $1+\frac{F_{1}^{s}}{d}\rightarrow 0$%
. We shall show that the effect of scattering in this channel is equivalent
to results obtained from $U(1)$ gauge theory.

\begin{figure}[htbp]
\includegraphics[width=4.8cm]{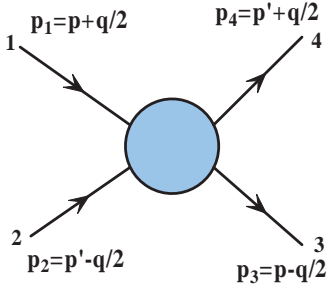}
\caption{(Color Online) Two quasi-particle scattering in a Fermi liquid.
Two in-going quasiparticles with momenta $\mathrm{p_1}$ and $\mathrm{p_2}$ interact with each other,
resulting in two out-going quasiparticles with momenta $\mathrm{p_3}$ and $\mathrm{p_4}$.
The momentum conservation requires that $\mathrm{p_1}+\mathrm{p_2}=\mathrm{p_3}+\mathrm{p_4}$.
The momentum transfer is $\mathrm{q}=\mathrm{p}_{1}-\mathrm{p}_{3}=\mathrm{p}_{4}-\mathrm{p}_{2}$.
By introducing $\mathrm{p}=\frac{1}{2}(\mathrm{p}_{1}+\mathrm{p}_{3})$ and 
$\mathrm{p}^{\prime }=\frac{1}{2}(\mathrm{p}_{2}+\mathrm{p}_{4})$, the four momenta $\mathrm{p_1}$, $\mathrm{p_2}$,
$\mathrm{p_3}$ and $\mathrm{p_4}$ can be written in terms of $\mathrm{p}$, $\mathrm{p}^{\prime }$ and $\mathrm{q}$.
}
\label{scattering}
\end{figure}

In the Landau transport equation which is essentially a Boltzmann equation,
the transition probability $W(1,2;3,4)$ for a two quasi-particles scattering
process in an isotropic Fermi liquid, $1+2\rightarrow 3+4$ with $i\equiv (%
\mathrm{p}_{i},\sigma _{i})$, is given by $2\pi $ times the squared moduli
of the quasi-particle scattering amplitude,%
\begin{equation*}
W(1,2;3,4)=2\pi |A(1,2;3,4)|^{2}.
\end{equation*}%
We are interested in the situation that the momentum transfer $\mathrm{q}=%
\mathrm{p}_{1}-\mathrm{p}_{3}$ is small and $\mathrm{p}=\frac{1}{2}(\mathrm{p%
}_{1}+\mathrm{p}_{3})$ and $\mathrm{p}^{\prime }=\frac{1}{2}(\mathrm{p}_{2}+%
\mathrm{p}_{4})$ are close to the Fermi momentum $\mathrm{p}_{F}$ as shown
in Fig. \ref{scattering}. In this case, the scattering amplitude depends
mainly on the relative orientation of the vector $\mathrm{p}$, $\mathrm{p}%
^{\prime }$ and $\mathrm{q}$, and on the the energy transfer $\omega
=\epsilon _{\mathrm{p+q/2}}-\epsilon _{\mathrm{p-q/2}}=\epsilon _{\mathrm{p}%
^{\prime }\mathrm{+q/2}}-\epsilon _{\mathrm{p}^{\prime }\mathrm{-q/2}}$. The
transition probability can be written as%
\begin{equation*}
W(1,2;3,4)=2\pi |A_{\mathrm{pp}^{\prime }}(\mathrm{q},\omega =\epsilon _{%
\mathrm{p+q/2}}-\epsilon _{\mathrm{p-q/2}})|^{2}.
\end{equation*}

\subsection{Quasi-particle scattering amplitude}

We shall neglect the spin indices in the following for brevity. The
spin-degeneracy factor 2 will be inserted when the need arises. The
quasi-particle scattering amplitude $A_{\mathrm{pp}^{\prime }}\left( \mathrm{%
q},\omega \right) $ is then given by the following equation,%
\begin{equation*}
A_{\mathrm{pp}^{\prime }}\left( \mathrm{q},\omega \right) -\sum_{\mathrm{p}%
^{\prime \prime }}f_{\mathrm{pp}^{\prime \prime }}\chi _{0\mathrm{p}^{\prime
\prime }}\left( \mathrm{q},\omega \right) A_{\mathrm{p}^{\prime \prime }%
\mathrm{p}^{\prime }}\left( \mathrm{q},\omega \right) =f_{\mathrm{pp}%
^{\prime }},
\end{equation*}%
where $\chi _{0\mathrm{p}}\left( \mathrm{q},\omega \right) $ is the
susceptibility%
\begin{equation*}
\chi _{0\mathrm{p}}\left( \mathrm{q},\omega \right) =\frac{n_{\mathrm{p}-%
\mathrm{q}/2}^{0}-n_{\mathrm{p}+\mathrm{q}/2}^{0}}{\omega +\xi _{\mathrm{p}-%
\mathrm{q}/2}-\xi _{\mathrm{p}+\mathrm{q}/2}}\simeq \frac{\mathbf{q}\cdot 
\mathbf{v}_{\mathrm{p}}}{\mathrm{q}\cdot \mathrm{v}_{\mathrm{p}}-\omega }%
\frac{\partial n_{\mathrm{p}}^{0}}{\partial \xi _{\mathrm{p}}},
\end{equation*}%
with $n_{\mathrm{k}}^{0}=n_{F}(\xi _{\mathrm{k}})$. We shall assume that the
scattering is dominating by the $l=1$ channel and approximate 
\begin{equation*}
f_{\mathrm{pp}^{\prime }}\sim \frac{\mathbf{p}\cdot \mathbf{p}^{\prime }}{%
p_{F}^{2}}f_{1}^{s}.
\end{equation*}%
It is then easy to show that%
\begin{equation}
A_{\mathrm{pp}^{\prime }}\left( \mathrm{q},\omega \right) =\frac{\mathbf{p}%
\cdot \mathbf{p}^{\prime }}{p_{F}^{2}}\frac{f_{1}^{s}}{1-\chi _{1}(\mathrm{q}%
,\omega )f_{1}^{s}},  \label{App0}
\end{equation}%
where%
\begin{equation*}
\chi _{1}(\mathrm{q},\omega )=\frac{1}{V}\sum_{\mathrm{p}}\frac{p^{2}}{%
p_{F}^{2}d}\left( \frac{n_{\mathrm{p}-\mathrm{q}/2}^{0}-n_{\mathrm{p}+%
\mathrm{q}/2}^{0}}{\omega +\xi _{\mathrm{p}-\mathrm{q}/2}-\xi _{\mathrm{p}+%
\mathrm{q}/2}+i\delta }\right) .
\end{equation*}%
For small $\left( \mathrm{q},\omega \right) $ we have%
\begin{equation*}
\chi _{1}(\mathrm{q},\omega )\sim -\frac{N(0)}{d}\left[ 1+ig(d)\frac{\omega 
}{v_{F}q}\right]
\end{equation*}%
for $\omega \ll v_{F}q$, where $q=|\mathrm{q}|$, $g(2)=1$ and $g(3)=\frac{%
\pi }{2}$. In the limit $N(0)f_{1}^{s}/d=F_{1}^{s}/d\rightarrow -1$, using
the expansion (see Eq.(\ref{f1})) 
\begin{equation*}
f_{1}^{s}=\frac{d}{N(0)}\left( -1-\beta \omega ^{2}+\gamma _{t}q^{2}\right) ,
\end{equation*}%
we obtain%
\begin{equation}
A_{\mathrm{pp}^{\prime }}\left( \mathrm{q},\omega \right) \simeq \frac{d}{%
N(0)}\frac{\mathbf{p}\cdot \mathbf{p}^{\prime }}{p_{F}^{2}}\frac{1}{-ig\frac{%
\omega }{v_{F}q}+\gamma _{t}q^{2}},  \label{App1}
\end{equation}%
where the last factor is exact the gauge field propagator in $U(1)$ gauge
theory.

\subsection{Thermal conductivity}

Following Pethick\cite{Pethick}, we use a variational approach\cite{Ziman}
to derive the thermal conductivity $\kappa $ from the transport equation.
The thermal resistivity for a Fermi liquid is given by%
\begin{eqnarray}
\frac{1}{\kappa } &=&\frac{1}{4}%
\sum_{1,2,3,4}W(1,2;3,4)n_{1}^{0}n_{2}^{0}(1-n_{3}^{0})(1-n_{4}^{0})  \notag
\\
&&\times (\phi _{1}+\phi _{2}-\phi _{3}-\phi _{4})^{2}\left( \sum_{1}\phi
_{1}\xi _{1}\mathbf{v}_{1}\cdot \mathbf{u}\frac{\partial n_{1}^{0}}{\partial
\epsilon _{1}}\right) ^{-2}  \notag \\
&&\times \delta (\epsilon _{1}+\epsilon _{2}-\epsilon _{3}-\epsilon
_{4})\delta _{\sigma _{1}+\sigma _{2},\sigma _{3}+\sigma _{4}}\delta _{%
\mathrm{p}_{1}+\mathrm{p}_{2},\mathrm{p}_{3}+\mathrm{p}_{4}},  \label{kappa1}
\end{eqnarray}%
where $n_{i}^{0}=n_{F}(\xi _{i})$ is the Fermi distribution function with $%
i=1,2,3,4$, $\xi _{i}=\epsilon_{i}-\mu$, $\phi _{i}$ is defined by $%
n_{i}=n_{i}^{0}-\phi _{i}\frac{\partial n_{i}^{0}}{\partial \epsilon _{i}}$, 
$\mathrm{v}_{i}$ is the quasiparticle velocity, \textrm{u} is an arbitrary
unit vector along which the temperature gradient $\nabla T$ is applied.

We shall make the usuall approximation $\phi _{i}=\xi _{i}\mathbf{v}%
_{i}\cdot \mathbf{u}$. To the order we are working with the approximation
that the quasi-particle velocity may be replaced by $v_{F}$ and%
\begin{eqnarray*}
\sum_{1}\phi _{1}\xi _{1}\mathbf{v}_{1}\cdot \mathbf{u}\frac{\partial
n_{1}^{0}}{\partial \epsilon _{1}} &=&\sum_{1}\left( \xi _{1}\mathbf{v}%
_{1}\cdot \mathbf{u}\right) ^{2}\frac{\partial n_{1}^{0}}{\partial \epsilon
_{1}} \\
&=&\frac{4m^{\ast }N(0)}{d}\int d\xi (\xi +\mu )\xi ^{2}\frac{\partial
n^{0}(\xi )}{\partial \xi } \\
&=&-\frac{4m^{\ast }N(0)}{d}\frac{\pi }{3}\epsilon _{F}(k_{B}T)^{2} \\
&=&-\frac{2\pi }{3}\frac{n}{m^{\ast }}(k_{B}T)^{2},
\end{eqnarray*}%
where $n$ is the fermion density and the relation $d(n/m^{\ast
})=N(0)v_{F}^{2}$ is used.

Introducing $\bar{\xi}_{\mathrm{p}}=\frac{1}{2}(\xi _{\mathrm{p+q/2}}+\xi _{%
\mathrm{p-q/2}})$ and using the conditions $\mathrm{q}=\mathrm{p}_{1}-%
\mathrm{p}_{3}=\mathrm{p}_{4}-\mathrm{p}_{2}$ and $\omega =\epsilon _{%
\mathrm{p+q/2}}-\epsilon _{\mathrm{p-q/2}}=\epsilon _{\mathrm{p}^{\prime }%
\mathrm{+q/2}}-\epsilon _{\mathrm{p}^{\prime }\mathrm{-q/2}}$, we have%
\begin{eqnarray*}
&&m^{\ast }\left( \xi _{1}\mathbf{v}_{1}+\xi _{2}\mathbf{v}_{2}-\xi _{3}%
\mathbf{v}_{3}-\xi _{4}\mathbf{v}_{4}\right) \\
&=&\left( \bar{\xi}_{\mathrm{p}}+\omega /2\right) \left( \mathbf{p}+\mathbf{q%
}/2\right) +\left( \bar{\xi}_{\mathrm{p}^{\prime }}-\omega /2\right) \left( 
\mathbf{p}^{\prime }-\mathbf{q}/2\right) \\
&&-\left( \bar{\xi}_{\mathrm{p}}-\omega /2\right) \left( \mathbf{p}-\mathbf{q%
}/2\right) -\left( \bar{\xi}_{\mathrm{p}^{\prime }}+\omega /2\right) \left( 
\mathbf{p}^{\prime }+\mathbf{q}/2\right) \\
&=&\left( \bar{\xi}_{\mathrm{p}}-\bar{\xi}_{\mathrm{p}^{\prime }}\right) 
\mathbf{q+}\omega \left( \mathbf{p}-\mathbf{p}^{\prime }\right) \\
&=&\left( \xi _{\mathrm{p}}-\xi _{\mathrm{p}^{\prime }}\right) \mathbf{q+}%
\omega \left( \mathbf{p}-\mathbf{p}^{\prime }\right) ,
\end{eqnarray*}%
and%
\begin{eqnarray*}
&&\langle \left( \phi _{1}+\phi _{2}-\phi _{3}-\phi _{4}\right) ^{2}\rangle
\\
&=&\langle \left[ \left( \xi _{1}\mathbf{v}_{1}+\xi _{2}\mathbf{v}_{2}-\xi
_{3}\mathbf{v}_{3}-\xi _{4}\mathbf{v}_{4}\right) \cdot \mathbf{u}\right]
^{2}\rangle \\
&=&\frac{1}{d}\left( \xi _{1}\mathbf{v}_{1}+\xi _{2}\mathbf{v}_{2}-\xi _{3}%
\mathbf{v}_{3}-\xi _{4}\mathbf{v}_{4}\right) ^{2} \\
&=&\frac{1}{m^{\ast 2}d}\left[ (\xi _{\mathrm{p}}-\xi _{\mathrm{p}^{\prime
}})\mathbf{q+}(\mathbf{p}-\mathbf{p}^{\prime })\omega \right] ^{2},
\end{eqnarray*}%
where $\langle \cdots \rangle $ means averaging over different \textrm{u}.

Putting $A_{\mathrm{pp}^{\prime }}\left( \mathrm{q},\omega \right) $ into (%
\ref{kappa1}), and using the identity $%
n_{1}^{0}(1-n_{3}^{0})=(n_{1}^{0}-n_{3}^{0})/[1-e^{\beta (\epsilon
_{1}-\epsilon _{3})}]$, we obtain

\begin{eqnarray*}
\frac{1}{\kappa } &\propto &\frac{1}{T^{4}}\sum_{\mathrm{q,p,p}^{\prime
}}\int d\omega \frac{|A_{\mathrm{pp}^{\prime }}\left( \mathrm{q},\omega
\right) |^{2}}{(e^{\beta \omega }-1)(1-e^{-\beta \omega })} \\
&&\times \lbrack n_{F}(\xi _{\mathrm{p-q/2}})-n_{F}(\xi _{\mathrm{p+q/2}%
})]\delta (\omega -\mathbf{p}\cdot \mathbf{q}/m^{\ast }) \\
&&\times \lbrack n_{F}(\xi _{\mathrm{p}^{\prime }\mathrm{+q/2}})-n_{F}(\xi _{%
\mathrm{p}^{\prime }\mathrm{-q/2}})]\delta (\omega -\mathbf{p}^{\prime
}\cdot \mathbf{q}/m^{\ast }) \\
&&\times \lbrack (\xi _{\mathrm{p}}-\xi _{\mathrm{p}^{\prime }})^{2}q^{2}%
\mathbf{+(p}-\mathbf{p}^{\prime })^{2}\omega ^{2}],
\end{eqnarray*}%
where we have used the $\delta$-functions to simplify the expression.
Replacing $n_{F}(\xi _{\mathrm{p-q/2}})-n_{F}(\xi _{\mathrm{p+q/2}})$ and $%
n_{F}(\xi _{\mathrm{p}^{\prime }\mathrm{+q/2}})-n_{F}(\xi _{\mathrm{p}%
^{\prime }\mathrm{-q/2}})$ by $\omega \frac{\partial n_{F}}{\partial \xi _{%
\mathrm{p}}}$ and $\omega \frac{\partial n_{F}}{\partial \xi _{\mathrm{p}%
^{\prime }}}$ respectively which is valid at small $\mathrm{q}$ and $\omega$%
, we obtain

\begin{eqnarray*}
\frac{1}{\kappa } &\propto &\frac{1}{T^{4}}\sum_{\mathrm{q,p,p}^{\prime
}}\int d\omega \frac{\omega ^{2}|A_{\mathrm{pp}^{\prime }}\left( \mathrm{q}%
,\omega \right) |^{2}}{(e^{\beta \omega }-1)(1-e^{-\beta \omega })} \\
&&\times \frac{\partial n_{F}}{\partial \xi _{\mathrm{p}}}\frac{\partial
n_{F}}{\partial \xi _{\mathrm{p}^{\prime }}}\delta (\omega -\mathbf{p}\cdot 
\mathbf{q}/m^{\ast })\delta (\omega -\mathbf{p}^{\prime }\cdot \mathbf{q}%
/m^{\ast }) \\
&&\times \lbrack (\xi _{\mathrm{p}}-\xi _{\mathrm{p}^{\prime }})^{2}q^{2}%
\mathbf{+(p}-\mathbf{p}^{\prime })^{2}\omega ^{2}].
\end{eqnarray*}%
Let $\theta (\theta ^{\prime })$ be the angle between $\mathbf{p}(\mathbf{p}%
^{\prime })$ and $\mathbf{q}$, and integrating over $\xi _{\mathrm{p}}$ and $%
\xi _{\mathrm{p}^{\prime }}$, we obtain 
\begin{eqnarray*}
\frac{1}{\kappa } &\propto &\frac{1}{T^{4}}\sum_{\mathrm{q,\hat{p},\hat{p}}%
^{\prime }}\int d\omega \frac{\omega ^{4}|A_{\mathrm{pp}^{\prime }}\left( 
\mathrm{q},\omega \right) |^{2}}{(e^{\beta \omega }-1)(1-e^{-\beta \omega })}%
\mathbf{(\hat{p}}-\mathbf{\hat{p}}^{\prime })^{2} \\
&&\times \delta (\omega -qv_{F}\cos \theta )\delta (\omega -qv_{F}\cos
\theta ^{\prime }).
\end{eqnarray*}%
Assuming that the scattering is dominated by $F_{1}^{s}$ channel and using
Eq.(\ref{App1}), we obtain 
\begin{eqnarray*}
\frac{1}{\kappa } &\propto &\frac{1}{T^{4}}\sum_{\mathrm{q}}\int d\omega 
\frac{\omega ^{4}}{(e^{\beta \omega }-1)(1-e^{-\beta \omega })}\frac{1}{g^{2}%
\frac{\omega ^{2}}{v_{F}^{2}q^{2}}+\gamma _{t}^{2}q^{4}} \\
&\propto &\frac{1}{T^{4}}\int d\omega \frac{\omega ^{4}}{(e^{\beta \omega
}-1)(1-e^{-\beta \omega })}\omega ^{-(4-d)/3} \\
&\propto &\left( \frac{k_{B}T}{\epsilon _{F}}\right) ^{(d-1)/3},
\end{eqnarray*}%
$\epsilon _{F}=p_{F}^{2}/2m^{\ast }$ is the spinon Fermi energy. The
expression represents the thermal resistivity coming from inelastic
scattering between fermions. At low temperature the inelastic scattering is
cut off by elastic impurity scattering rate $1/\tau _{0}$, which gives rise
to 
\begin{equation*}
\frac{\kappa _{\text{el}}}{T}=\frac{1}{d}\gamma ^{\ast }v_{F}^{2}\tau _{0},
\end{equation*}%
where $\gamma ^{\ast }=C_{V}/T$ is the specific heat ratio, and $v_{F}$ is
the spinon Fermi velocity. The total thermal conductivity is therefore given
by%
\begin{equation*}
\frac{\kappa }{T}\propto \max \left[ \frac{\hbar }{k_{B}^{2}}\left( \frac{%
k_{B}T}{\epsilon _{F}}\right) ^{(4-d)/3},\frac{d}{\gamma ^{\ast }v_{F}^{2}}%
\frac{1}{\tau _{0}}\right] ^{-1}.
\end{equation*}%
The same result is obtained in U(1) gauge theory at 2D.\cite{Nave07}

\section{Discussions}

\subsection{Relation to the $U(1)$ gauge theory}

The U(1) spin liquid is actually a member of the QSLs described by our
phenomenology keeping scattering in the $l=0,1$ channels only. To show this
we start with a Landau Fermi liquid with interaction parameters $F_{0}^{s}$
and $F_{1}^{s}(q,\omega )$ only. The long-wavelength and low dynamics of the
Fermi liquid is described by an effective Lagrangian 
\begin{equation}
L_{\text{eff}}=\sum_{\mathrm{k},\sigma }\left[ c_{\mathrm{k}\sigma
}^{\dagger }(i\frac{\partial }{\partial t}-\xi _{\mathrm{k}})c_{\mathrm{k}%
\sigma }-H^{\prime }(c^{\dag },c)\right] ,  \label{leff}
\end{equation}%
where $c_{\mathrm{k}\sigma }^{\dagger }(c_{\mathrm{k}\sigma })$ are spin-$%
\sigma $ fermion creation (annihilation) operators with momentum $\mathrm{k}$%
, and 
\begin{equation}
H^{\prime }(c^{\dag },c)=\frac{1}{2N(0)}\sum_{q}\left[ \frac{F_{1}^{s}}{%
v_{F}^{2}}\mathbf{j}(q)\cdot \mathbf{j}(-q)+F_{0}^{s}n(q)n(-q)\right]
\label{h'eff}
\end{equation}%
describes the current-current and density-density interactions between
quasi-particles\cite{Leggett}, where $q=(\mathrm{q},\omega )$ and $%
v_{F}=\hbar k_{F}/m^{\ast }$ is the Fermi velocity.

The current- and density- interactions can be decoupled by introducing
fictitious gauge potentials $\mathbf{a}$ and $\varphi $
(Hubbard-Stratonovich transformation) with 
\begin{equation}
H^{\prime }(c^{\dag },c)\rightarrow \sum_{q}\left[ \mathbf{j}\cdot \mathbf{a}%
+n\varphi -{\frac{1}{2}}\left( \frac{n}{m^{\ast }}\frac{d}{F_{1}^{s}}\mathbf{%
a}^{2}+\frac{N(0)}{F_{0}^{s}}\varphi ^{2}\right) \right] ,  \label{h'eff2}
\end{equation}%
where $n$ is fermion density. We have used the equality $d(n/m^{\ast
})=N(0)v_{F}^{2}$ in writing down Eq.\ (\ref{h'eff2}).

The Lagrangian\ (\ref{leff}) and\ (\ref{h'eff2}) can be rewritten in the
standard form of $U(1)$ gauge theory by noting that the fermion current is
given in this representation by 
\begin{equation*}
\mathbf{j}=\frac{-i}{2m^{\ast }}\sum_{\sigma }\left[ \psi _{\sigma }^{\dag
}\nabla \psi _{\sigma }-(\nabla \psi _{\sigma }^{\dagger })\psi _{\sigma }%
\right] -\frac{n}{m^{\ast }}\mathbf{a},
\end{equation*}%
where $\psi _{\sigma }(\mathrm{r})=\int e^{-i\mathrm{k\cdot r}}c_{\mathrm{k}%
\sigma }$ in the Fourier transform of $c_{\mathrm{k}\sigma }$. The
Lagrangian can be written as 
\begin{subequations}
\label{lspin}
\begin{equation}
L=\sum_{\sigma }\int d^{d}\mathrm{r}\left[ \psi _{\sigma }^{\dagger }(i\frac{%
\partial }{\partial t}-\varphi )\psi _{\sigma }-H(\psi _{\sigma }^{\dagger
},\psi _{\sigma })\right] +L(\varphi ,\mathbf{a}),  \label{leff2}
\end{equation}%
where 
\begin{equation}
H(\psi _{\sigma }^{\dagger },\psi _{\sigma })=\frac{1}{2m^{\ast }}|(\nabla -i%
\mathbf{a})\psi _{\sigma }|^{2}  \label{heff}
\end{equation}%
and 
\begin{equation}
L(\varphi ,\mathbf{a})={\frac{1}{2}}\int d^{d}\mathrm{r}\left[ \frac{n}{%
m^{\ast }}(1+\frac{d}{F_{1}^{s}})\mathbf{a}^{2}+\frac{N(0)}{F_{0}^{s}}%
\varphi ^{2}\right] .
\end{equation}%
Notice how the ${\frac{n}{2m^{\ast }}}\mathbf{a}^2$ term in $L(\varphi ,%
\mathbf{a})$ arises from the introduction of diamagnetic term in $H(\psi
_{\sigma }^{\dag },\psi _{\sigma })$.

Using Eq.\ (\ref{f1}), we find that in the small $q$ limit, the transverse
part of $L(\varphi ,\mathbf{a})$ is given in the spin liquid state by 
\end{subequations}
\begin{equation}
L_{t}(\varphi ,\mathbf{a})=-\frac{n}{2m^{\ast }}\int d^{d}\mathrm{r}\left[
\beta (\frac{\partial \mathbf{a}}{\partial t})^{2}-\gamma _{t}(\nabla \times 
\mathbf{a})^{2}\right] .  \label{lt}
\end{equation}%
The longitudinal part of the gauge potential $\varphi $ is screened as long
as $F_{0}^{s}\neq 0$. Lagrangian (\ref{lspin}) together with (\ref{lt}) is
the standard Lagrangian used to describe $U(1)$ QSLs. It is interesting to
note that a nonzero $1+\frac{F_{1}^{s}(0,0)}{d}$ leads to a mass term for
the gauge field $\mathbf{a}$, in agreement with slave-boson/rotor approaches
where a metallic state appears with condensation of bosons/rotors\cite%
{Lee05,Ng1}.

\subsection{Alternative picture of Mott transition}

The close relation between Fermi liquid and spin-liquid states suggests that
the (zero temperature) metal-insulator transition between the two states is
characterized by the change of Landau parameter $1+F_{1}^{s}(0,0)/d%
\rightarrow 0^{+}$ across the transition. The nature of metal-insulator
transition within the Fermi liquid framework was first addressed by Brinkman
and Rice\cite{BrinkmanRice} where they proposed that a metal-insulator
(Mott) transition is indicated by diverging effective mass ${\frac{m^{\ast }%
}{m}}\rightarrow \infty $ and inverse compressibility $\kappa \rightarrow 0$
at the Mott transition point, with correspondingly a vanishing
quasi-particle renormalization weight $Z\sim {\frac{m}{m^{\ast }}}%
\rightarrow 0$. The diverging effective mass and vanishing quasi-particle
weight suggest that the Fermi liquid state is destroyed at the Mott
transition, and the Mott insulator state is distinct from the Fermi liquid
state at the metal side.

\begin{figure}[tbph]
\includegraphics[width=6.4cm]{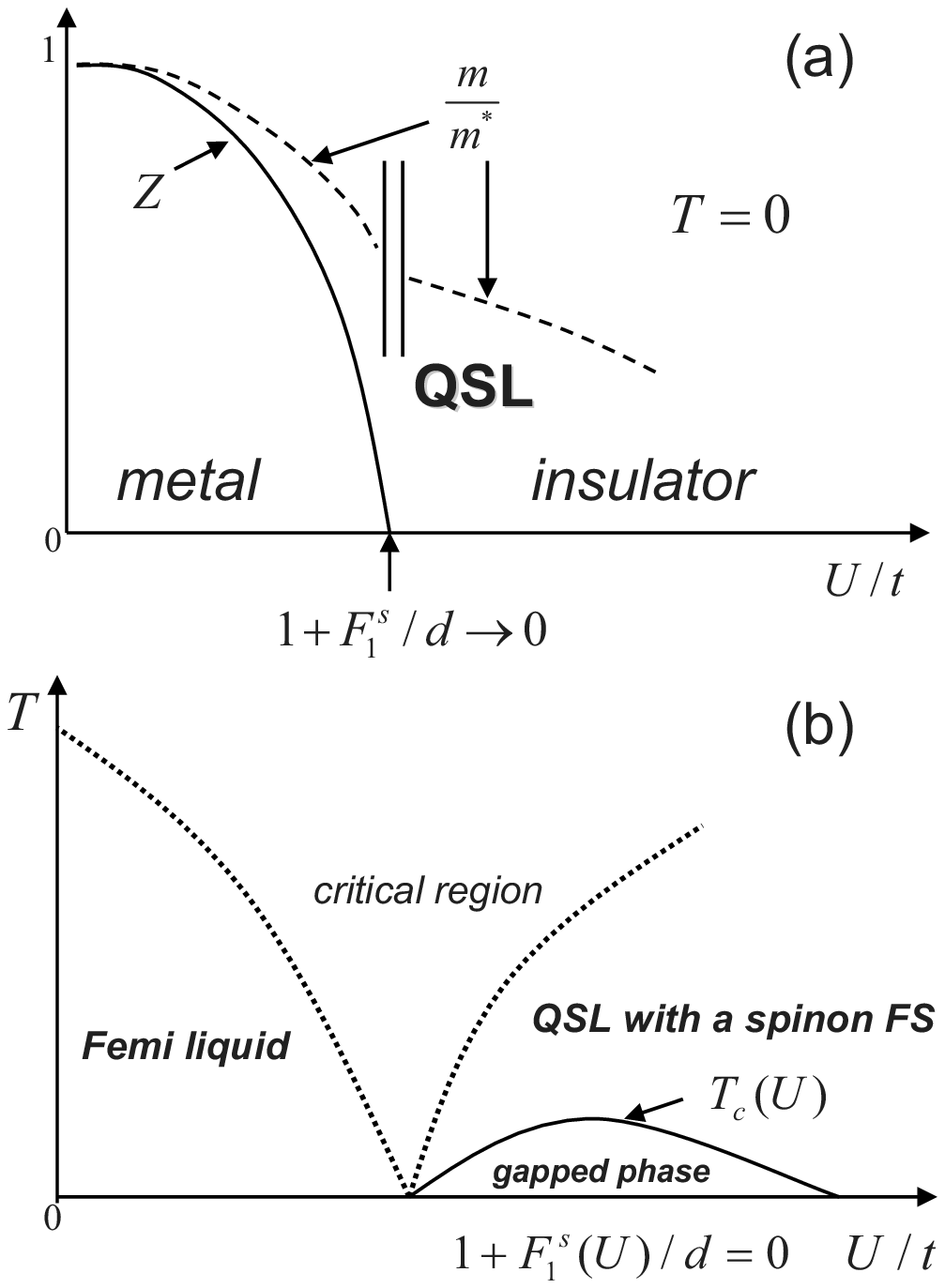}
\caption{ (a) Schematic zero temperature phase diagram for Mott transition. $%
U$ is the Hubbard interaction strength and $t$ is the hopping integral. The
electron quasiparticle weight and quasi-particle charge current $\sim
1+F_1^s/d$ vanishes at the critical point while the effective mass remains
finite. (b) Schematic phase diagram showing finite temperature crossovers
and possible instability toward gapped phases at lower temperature. There
exists a (finite temperature) critical region around $U_c$ where our
phenomenological theory is not applicable.}
\label{phases}
\end{figure}

Here we propose an alternative picture where the Fermi surface is not
destroyed but the quasiparticles are converted into spinons at the Mott
transition. In particular, the effective mass $m^{\ast }/m$ may not diverge
at the metal-insulator transition although $Z\rightarrow 0$ in this picture.
A schematic phase diagram for the Mott (metal-QSLs) transition is presented
in Fig.\ (\ref{phases}) where we imagine a Hubbard type Hamiltonian with
hopping $t$ and on-site Coulomb repulsion $U$. The system is driven to a
Mott insulator state at zero temperature at $U=U_{c}$, where $%
1+F_{1}^{s}(U>U_{c})/d=0$. Our picture is supported by the experimental fact
that the potential candidates for the $U(1)$ QSLs with large Fermi surfaces
(ET and dmit salts) are all closed to the metal-insulator transition. We
caution that in general a finite ($T\neq 0$) region exists around the Mott
transition point where the physics is dominated by critical fluctuations and
our phenomenological theory is not applicable. We note that an alternative
phenomenology for Mott transition from a semi-microscopic starting point\cite%
{Senthil} has qualitatively similar conclusion as our present work. The
relation between the two works is not clear at present.

\subsection{Pomeranchuk instability}

Experienced researchers in Fermi liquid theory will recognize that the point 
$1+F_{1}^{s}/d=0$ is in fact a critical point in Fermi liquid theory. The
Fermi surface is unstable with respect to deformation when $1+F_{1}^{s}/d<0$%
. The stability of the $1+F_{1}^{s}/d=0$ point is required in QSLs where
quasi-particles (spinons) become chargeless. The resulting QSLs we obtain
here are marginally stable because of large critical fluctuations. The large
critical fluctuations give rise to singular corrections to thermodynamics
quantities (specific heat for example) and transport coefficients (various
scattering lifetimes) at two dimensions as first pointed out in $U(1)$ gauge
theory. The Pomeranchuk criticality is an alternative way to express these
results.

The presence of Pomeranchuk criticality suggests that QSLs with large Fermi
surfaces are in general rather susceptible to formation of other more stable
QSLs at lower temperature, like the $Z_{2}$ QSLs or valence bond solid (VBS)
states that gap out part or the whole Fermi surface. The resulting phase
diagram at the vicinity of Mott transition thus has the generic feature
shown in Fig.(\ref{phases}b), where the system is driven into a gapped QSL
at low temperature $T<T_{c}(U)$ at the insulating side. The nature of the
low temperature QSLs depends on the microscopic details of the system and
cannot be determined from our phenomenology. Our theory is applicable at $%
T>T_{c}(U)$, when the spin liquid is still in the large-Fermi surface phase.

\section{Conclusion}

In summary, we formulate a Fermi liquid type phenomenological theory for
quantum spin liquid states in the vicinity of metal-insulator transition.
The phenomenology takes into account the fact that DC electrical current and
compressibility vanishes while thermal current keeps finite in QSLs.
Physically, the phenomenology implies that the charge degrees of freedom of
quasiparticles are frozen at $\mathrm{q}=0$ and $\omega =0$. Finite specific
heat ratio, spin susceptibility and the fact Wilson ration is of order of
unity indicate spin degrees of freedom are still active in this limit. The
frozen charge degrees of freedom are recovered at finite \textrm{q} and $%
\omega $ as indicated by the power-law $\omega^{\eta}$-dependent AC
conductivity. We also show that the $U(1)$ spin liquids is a member of the
class of QSLs described by our phenomenology. We also propose an alternative
picture of Mott transition and discuss the phase diagram.

TKN is supported by HKRGC through grant HKUST03/CRF09 and GRF 603410. YZ is
supported by National Basic Research Program of China (973 Program,
No.2011CBA00103/2014CB921201), NSFC (No.11074218/11374256) and the
Fundamental Research Funds for the Central Universities in China. He also
thanks the Institute for Advanced Studies, HKUST for its hospitality, where
this work is completed.

\appendix

\section{Renormalized currents}

In this appendix we derive the renormalized currents in Fermi liquid theory
((Eq.(\ref{Je}) and (\ref{JQ})) in the main text). The local equilibrium
quasi-particle occupation numbers and their fluctuations have to be
considered carefully. The excitation energy of an additional quasiparticle
with momentum $\mathrm{p}$ is given by%
\begin{equation*}
\tilde{\epsilon}_{\mathrm{p}}=\epsilon _{\mathrm{p}}+\sum_{\mathrm{p}%
^{\prime }}f_{\mathrm{pp}^{\prime }}^{s}\delta n_{\mathrm{p}^{\prime }},
\end{equation*}%
where $\epsilon _{\mathrm{p}}=\frac{p^{2}}{2m^{\ast }}$. The corresponding
local equilibrium occupation number is $\tilde{n}_{\mathrm{p}}^{0}\equiv
n_{F}\left( \tilde{\epsilon}_{\mathrm{p}}-\mu \right) $, and the departure
from local equilibrium reads%
\begin{eqnarray*}
\delta \tilde{n}_{\mathrm{p}} &=&n_{\mathrm{p}}-\tilde{n}_{\mathrm{p}}^{0} \\
&=&\delta n_{\mathrm{p}}-\frac{\partial n^{0}}{\partial \epsilon _{\mathrm{p}%
}}\sum_{\mathrm{p}^{\prime }}f_{\mathrm{pp}^{\prime }}^{s}\delta n_{\mathrm{p%
}^{\prime }},
\end{eqnarray*}%
where $\delta n_{\mathrm{p}}=n_{\mathrm{p}}-n_{\mathrm{p}}^{0}$. The charge
current $\mathrm{J}$ carried by quasiparticles is related to the particle
density by the conservation law,%
\begin{equation*}
\frac{\partial \rho }{\partial t}+\nabla _{\mathrm{r}}\cdot \mathbf{J}=0.
\end{equation*}%
The density fluctuation $\delta \rho \left( \mathrm{r},t\right) $ should be
expressed in terms of the sum of $\delta \tilde{n}_{\mathrm{p}}\left( 
\mathrm{r},t\right) $ (i.e. fluctuation away from local equilibirum),%
\begin{equation*}
\delta \rho \left( \mathrm{r},t\right) =\sum_{\mathrm{p}}\delta \tilde{n}_{%
\mathrm{p}}\left( \mathrm{r},t\right)
\end{equation*}%
and 
\begin{equation*}
\frac{\partial }{\partial t}\delta \rho +\nabla _{\mathrm{r}}\cdot \sum_{%
\mathrm{p}}\delta \tilde{n}_{\mathrm{p}}\mathbf{v}_{\mathrm{p}}=0.
\end{equation*}%
Therefore, 
\begin{equation*}
\mathbf{J}=\sum_{\mathrm{p}}\delta \tilde{n}_{\mathrm{p}}\mathbf{v}_{\mathrm{%
p}}=\sum_{\mathrm{p}}\delta n_{\mathrm{p}}\mathbf{j}_{\mathrm{p}},
\end{equation*}%
where%
\begin{equation*}
\mathbf{j}_{\mathrm{p}}=\mathbf{v}_{\mathrm{p}}-\sum_{\mathrm{p}^{\prime
}}f_{\mathrm{pp}^{\prime }}^{s}\frac{\partial n^{0}}{\partial \epsilon _{%
\mathrm{p}^{\prime }}}\mathbf{v}_{\mathrm{p}^{\prime }}.
\end{equation*}%
Using the relation%
\begin{eqnarray*}
\sum_{\mathrm{p}^{\prime }}\frac{\partial n^{0}}{\partial \epsilon _{\mathrm{%
p}^{\prime }}}f_{\mathrm{pp}^{\prime }}\mathbf{v}_{\mathrm{p}^{\prime }}
&=&\sum_{\mathrm{p}^{\prime }}\frac{\partial n^{0}}{\partial \epsilon _{%
\mathrm{p}^{\prime }}}N_{F}^{-1}\sum_{l}F_{l}^{s}P_{l}\left( \cos \theta
\right) \mathbf{v}_{\mathrm{p}^{\prime }} \\
&=&\frac{1}{d}F_{1}^{s}\mathbf{v}_{\mathrm{p}}\int d\epsilon ^{\prime }\frac{%
\partial n^{0}}{\partial \epsilon ^{\prime }}=\frac{1}{d}F_{1}^{s}\mathbf{v}%
_{\mathrm{p}},
\end{eqnarray*}%
where $N_{F}=N\left( 0\right) $, we find that the renormalized charge
current is%
\begin{equation*}
\mathbf{J}=\frac{m}{m^{\ast }}(1+\frac{1}{d}F_{1}^{s})\mathbf{J}^{(0)},
\end{equation*}%
when $\mathbf{J}^{0}$ is the electric current in the absence of interaction.
Notice that $P_{l}\left( \cos \theta \right) $ is replaced by $\cos \left(
l\theta \right) $ at 2D.

Similarly the thermal (energy) current $\mathbf{J}_{Q}$ is given by%
\begin{eqnarray*}
\mathbf{J}_{Q} &=&\sum_{\mathrm{p}}\delta \tilde{n}_{\mathrm{p}}\left(
\epsilon _{\mathrm{p}}-\mu \right) \mathbf{v}_{\mathrm{p}} \\
&=&\sum_{\mathrm{p}}\left( \epsilon _{\mathrm{p}}-\mu \right) \mathbf{v}_{%
\mathrm{p}}(\delta n_{\mathrm{p}}-\sum_{\mathrm{p}^{\prime }}\frac{\partial
n^{0}}{\partial \epsilon _{\mathrm{p}}}f_{\mathrm{pp}^{\prime }}\delta n_{%
\mathrm{p}^{\prime }}) \\
&=&\sum_{\mathrm{p}}\delta n_{\mathrm{p}}[\left( \epsilon _{\mathrm{p}}-\mu
\right) \mathbf{v}_{\mathrm{p}}-\sum_{\mathrm{p}^{\prime }}\frac{\partial
n^{0}}{\partial \epsilon _{\mathrm{p}^{\prime }}}f_{\mathrm{pp}^{\prime
}}\left( \epsilon _{\mathrm{p}^{\prime }}-\mu \right) \mathbf{v}_{\mathrm{p}%
^{\prime }}].
\end{eqnarray*}%
Notice that%
\begin{eqnarray*}
&&\sum_{\mathrm{p}^{\prime }}\frac{\partial n^{0}}{\partial \epsilon _{%
\mathrm{p}^{\prime }}}f_{\mathrm{pp}^{\prime }}\left( \epsilon _{\mathrm{p}%
^{\prime }}-\mu \right) \mathbf{v}_{\mathrm{p}^{\prime }} \\
&=&\sum_{\mathrm{p}^{\prime }}\frac{\partial n^{0}}{\partial \epsilon _{%
\mathrm{p}^{\prime }}}N_{F}^{-1}\sum_{l}F_{l}^{s}P_{l}\left( \cos \theta
\right) \left( \epsilon _{\mathrm{p}^{\prime }}-\mu \right) \mathbf{v}_{%
\mathrm{p}^{\prime }} \\
&=&\frac{1}{d}F_{1}^{s}\mathbf{v}_{\mathrm{p}}\int d\epsilon ^{\prime }\frac{%
\partial n^{0}}{\partial \epsilon ^{\prime }}\left( \epsilon ^{\prime }-\mu
\right) =0
\end{eqnarray*}%
to leading order. Therefore the renormalized thermal current is given by%
\begin{equation*}
\mathbf{J}_{Q}=\frac{m}{m^{\ast }}\mathbf{J}_{Q}^{(0)}.
\end{equation*}%
We observe that the thermal current is not renormalized by the factor $%
\left( 1+\frac{1}{d}F_{1}^{s}\right) $.

\end{document}